\documentclass[12pt,fleqn]{article}
\usepackage{graphicx}

\def\be{\begin{equation}}
\def\ee{\end{equation}}
\def\bea{\begin{eqnarray}}
\def\eea{\end{eqnarray}}

\def\bbuildrel#1_#2^#3{\mathrel{\mathop{\kern 0pt#1}\limits_{#2}^{#3}}}
\def\slash#1{\setbox0=\hbox{$#1$}#1\hskip-\wd0\dimen0=5pt\advance
       \dimen0 by-\ht0\advance\dimen0 by\dp0\lower0.5\dimen0\hbox
         to\wd0{\hss\sl/\/\hss}}

\newcommand{\scs}{\scriptscriptstyle}
\newcommand{\f}{\frac}

\newcommand{\al}{\alpha_s}

\newcommand{\newsection}[1]{\section{#1}\setcounter{equation}{0}}

\begin{document}
\begin{titlepage}

\begin{flushright}
  {\bf TUM-HEP-468/02\\
       IFT-24/2002\\
       hep-ph/0207131}\\[2cm]  
\end{flushright}

\begin{center}

\setlength {\baselineskip}{0.3in} 
{\bf\Large $\bar{B} \to X_s \gamma$ after Completion of the NLO QCD Calculations}\\[2cm]

\setlength {\baselineskip}{0.2in}
{\large  Andrzej J. Buras$^{^{1}}$ and Miko{\l}aj Misiak$^{^{2}}$}\\[5mm]

$^{^{1}}${\it Physik Department, Technische Universit\"at M\"unchen,\\
               D-85748 Garching, Germany}\\[3mm]

$^{^{2}}${\it Institute of Theoretical Physics, Warsaw University,\\
                 Ho\.za 69, PL-00-681 Warsaw, Poland}\\[2cm] 

{\bf Abstract}\\
\end{center} 
\setlength{\baselineskip}{0.2in} 

Several years ago, Stefan Pokorski, Manfred M\"unz and us outlined a
program for calculation of the NLO QCD corrections to the weak
radiative $\bar{B}$ meson decay $\bar{B} \to X_s \gamma$. Very
recently, just before the 60th birthday of Stefan Pokorski, this
program has been formally completed.  In the present paper, we
summarize the existing results and discuss perspectives for further
improvement of the accuracy of the Standard Model prediction for ${\rm
BR}[\bar{B} \to X_s \gamma]$.\\

\begin{center}
{\bf  Submitted to the special issue of Acta Physica Polonica B\\
dedicated to Stefan Pokorski on the occasion of his 60th birthday.}
\end{center} 

\end{titlepage} 
\setlength{\baselineskip}{0.23in}

\newsection{Introduction}
\label{sec:intro}

\begin{figure}[h]
\begin{center}
\includegraphics[width=75mm,angle=0]{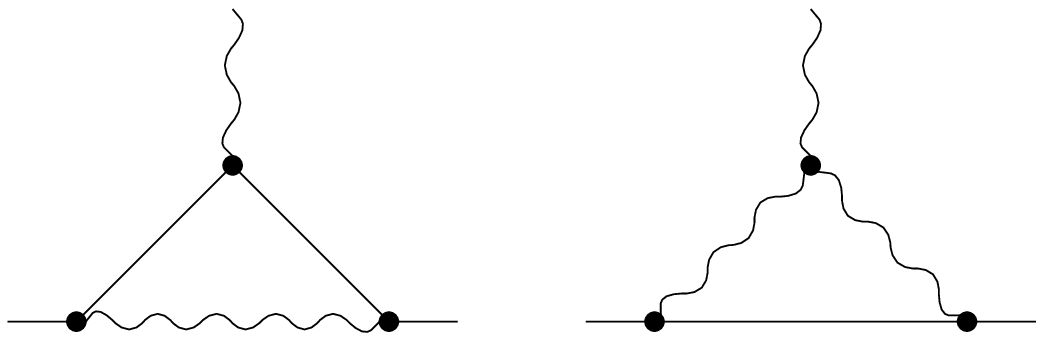}
\end{center}
\vspace*{-32mm}
\hspace*{59mm}  $\gamma$ \hspace{36.5mm} $\gamma$\\[1cm] 
\hspace*{47mm} $u,c,t$ \hspace{9mm}     $u,c,t$ \hspace{12mm} 
            $W^{\pm}$ \hspace{8mm}   $W^{\pm}$ \hspace{15mm}\\[5mm]
\hspace*{46mm} $b$ \hspace{1cm}  $W^{\pm}$  \hspace{7mm} $s$ \hspace{6mm} 
               $b$ \hspace{7mm}   $u,c,t$   \hspace{7mm} $s$\\[-1cm] 
\begin{center}
\caption{\sf Leading-order Feynman diagrams for $b \to s \gamma$ in the SM.}
\label{fig:SMdiag}
\end{center}
\vspace{-1cm}
\end{figure}

The radiative decay ${\bar B} \to X_s \gamma$ is known to be extremely
sensitive to the structure of fundamental interactions at the
electroweak scale. It is dominantly generated by the Flavour Changing
Neutral Current (FCNC) decay $b \to s \gamma$ that does not arise at
the tree level in the Standard Model (SM). The leading order SM
diagrams are shown in fig.~\ref{fig:SMdiag}.

Many possible non-standard contributions (e.g., SUSY one-loop
diagrams) are of the same order in electroweak interactions. They
might remain important even for relatively heavy exotic particles.
Consequently, $b \to s \gamma$ imposes severe constraints on
extensions of the SM (see, for instance,
\cite{Buras:xp}--\cite{Ellis:2001yu}).

The inclusive branching ratio BR$[{\bar B} \to X_s \gamma]$ has been
measured so far by CLEO \cite{Chen:2001fj}, \linebreak BELLE \cite{Abe:2001hk} and
ALEPH \cite{Barate:1998vz}. The most accurate result is the one of CLEO, where
photons with energies down to 2.0~GeV are included. Extrapolation
towards lower photon energy cutoffs is performed following the
phenomenological models of refs.~\cite{Ali:1990tj,Kagan:1998ym}.

When the photon energy cutoff is chosen to be 1.6~GeV in the
$\bar{B}$-meson rest frame, the experimental world average reads\footnote{
By convention, contributions to ${\bar B} \to X_s \gamma$ from
intermediate real $\psi$ and $\psi'$ are treated as background,
while all the other $c\bar{c}$ states are included.}
\be \label{exp}
{\rm BR}[\bar{B} \to X_s \gamma ~(E_\gamma > 1.6~{\rm GeV})]_{\rm exp}
 = (3.12 \pm 0.41) \times 10^{-4}.
\ee
Within $1\sigma$, it matches the SM prediction \cite{Gambino:2001ew,Buras:2002tp}
\be \label{thSM}
{\rm BR}[\bar{B} \to X_s \gamma ~ (E_\gamma > 1.6~{\rm GeV})]_{\rm SM} 
= (3.57 \pm 0.30) \times 10^{-4}.
\ee
One can see that the experimental and theoretical uncertainties are
close in size. Without the inclusion of the Next-to-Leading Order (NLO)
QCD corrections, the theoretical uncertainty in eq.~(\ref{thSM})
 would be around three times larger, and the constraints on new
physics --- much weaker.

The program of the NLO calculation was outlined by Stefan Pokorski,
Manfred M\"unz and us in the article \cite{Buras:xp}. At that time,
the only known results were the Leading Order (LO) ones that suffered
from large scale uncertainties~\cite{Ali:1993ct,Buras:xp}. We analyzed
these uncertainties in detail, and enumerated calculations that still
had to be done in the NLO case. Very recently, the last element of
this NLO program has been completed \cite{Buras:2002tp}. In parallel
to the QCD calculations, progress was being made in evaluation of the
electroweak corrections, non-perturbative effects, as well as in
collecting and analyzing the experimental data.

In the present paper, we summarize all the contributions to the NLO
QCD calculation of BR$[{\bar B} \to X_s \gamma]$, and discuss
perspectives for further improvement of the theoretical accuracy. In
particular, we point out the interplay between charm-quark mass
uncertainties in the perturbative calculation and non-perturbative
effects.

Our article is organized as follows. The next section is devoted to a
brief description of the history of perturbative calculations of QCD
effects in $b \to s \gamma$. In section~\ref{sec:ew}, we summarize the
electroweak corrections.  Non-perturbative effects are discussed in
section~\ref{sec:nonpert}. The main theoretical uncertainties and
possibilities for their elimination are the subject of
section~\ref{sec:phenom}. Section~\ref{sec:concl} contains our
conclusions.

\newsection{The LO and NLO QCD calculations}
\label{sec:perturb}

In a certain range of photon energy cutoffs, the width of the hadronic
decay $\bar{B} \to X_s \gamma$ is well approximated by the
perturbative decay width
\be \label{pert}
\Gamma[b \to X_s^{\rm parton} \gamma] = \Gamma[b \to s \gamma] + \Gamma[b \to s \gamma g] + \ldots.
\ee
Arguments that support such a statement will be discussed in section
4. Until then, we shall restrict our discussion to the perturbative
quantity (\ref{pert}).

The framework for all the renormalization-group-improved perturbative
analyses of \linebreak $b \to X_s^{\rm parton} \gamma$~ is set by the effective
Lagrangian
\be \label{Leff}
{\cal L_{\rm eff}} =  {\cal L}_{\scs {\rm QCD} \times {\rm QED}}(u,d,s,c,b) 
~+~ \f{4 G_F}{\sqrt{2}} V_{ts}^* V_{tb} \sum_{i=1}^8 C_i(\mu) Q_i ~+~ \ldots.
\ee
It is obtained from the underlying theory (SM in our case) by
decoupling of all the particles that are much heavier than the
$b$-quark. The Wilson coefficients $C_i(\mu)$ play the role of
coupling constants at the vertices $Q_i$. The generic structure of the
operators $Q_i$ is as follows:
\be \label{ops}
Q_i = \left\{ \begin{array}{ll}
(\bar{s} \Gamma_i c)(\bar{c} \Gamma'_{\underline{i}} b), & i=1,2, \\[3mm]
(\bar{s} \Gamma_i b) \sum_q (\bar{q} \Gamma'_{\underline{i}} q), 
& i=3,4,5,6,~~~~~~~~~ (q=u,d,s,c,b) \\[3mm]
\f{e m_b}{16 \pi^2} \bar{s}_L \sigma^{\mu \nu} b_R F_{\mu \nu}, & i=7, \\[3mm]
\f{g m_b}{16 \pi^2} \bar{s}_L \sigma^{\mu \nu} T^a b_R G^a_{\mu \nu},~~
& i=8. \end{array} \right.  
\ee
Here, $\Gamma_i$ and $\Gamma'_i$ denote various combinations of the colour
and Dirac matrices (see, e.g., \cite{Buras:2002tp}).

The dots in eq.~(\ref{Leff}) stand for UV-counterterms and
non-physical operators that vanish by the QCD$\times$QED equations of
motion. In the present section, we neglect everything that is
not important for $b \to s \gamma$ at the leading order in
$\alpha_{\rm em}$, $m_b/M_W$, $m_s/m_b$ and $V_{ub}/V_{cb}$. This
includes other operators $Q_i$ of dimension 5 and 6,
higher-dimensional operators, as well as terms involving leptons.

Let us assume that the decoupling of heavy particles is performed in
the $\overline{\rm MS}$ scheme, at the renormalization scale $\mu_0
\sim M_W$. The values of $C_i(\mu_0)$ are found from the so-called
matching conditions, i.e. by imposing equality of the effective- and
underlying-theory Green functions at external momenta that are much
smaller than masses of the decoupled particles. Next, the Wilson
coefficients are evolved from $\mu=\mu_0$ down to $\mu=\mu_b \sim
m_b$, according to the Renormalization Group Equations (RGE) 
\be \label{RGE}
\mu \f{d}{d \mu} C_i(\mu) = C_j(\mu) \gamma_{ji}(\mu),
\ee
where the anomalous dimension matrix $\hat{\gamma}$ is found from UV
divergences in the effective theory. This procedure results in
expressing the effective Lagrangian (\ref{Leff}) in terms of
\be \label{cmb}
C_i(\mu_b) = C_i^{(0)}(\mu_b)+ \f{\al(\mu_b)}{4\pi} C_i^{(1)}(\mu_b)
+ \left( \f{\al(\mu_b)}{4\pi} \right)^2 C_i^{(2)}(\mu_b) + \ldots,
\ee
where $C_i^{(n)}(\mu_b)$ depend on $\al$ only via the ratio $\eta
\equiv \al(\mu_0)/\al(\mu_b)$. Consequently, working at a fixed order
in $\al$, one truncates an expansion in powers of $\al(\mu_b)$ rather
than in powers of ~$\al(M_W) \ln(M_W^2/m_b^2)$,~ as it would be the
case without introduction of the effective theory. Thus, the behaviour
of the perturbation series improves. This is the essence of the
renormalization-group improvement in the considered case.

In the LO calculations, everything but $C_i^{(0)}(\mu_b)$ is neglected
in eq.~(\ref{cmb}). At the NLO, one takes into account all the ${\cal
O}(\al(\mu_b))$ contributions to $\Gamma[b \to X_s^{\rm parton}
\gamma]$, including those containing $C_i^{(1)}(\mu_b)$.

The Wilson coefficients encode information on the short-distance QCD
effects due to hard gluon exchanges between the quark lines of the
leading one-loop electroweak diagrams (fig.~\ref{fig:SMdiag}).  Such
effects enhance the branching ratio ${\rm BR}[{\bar B}\to X_s\gamma]$
by roughly a factor of three, as first pointed out in
refs.~\cite{Bertolini:1986th,Deshpande:nr}.  

A peculiar feature of the renormalization group analysis in $b\to
s\gamma$ is that the mixing under infinite renormalization between the
four-fermion operators $Q_1,...,Q_6$ and the ``magnetic penguin''
operators $Q_7,Q_8$, which govern this decay, vanishes at the
one-loop level. Consequently, in order to calculate the coefficients
$C_7(\mu_b)$ and $C_8(\mu_b)$ at LO, two-loop calculations are
necessary. Such calculations were completed in
ref.~\cite{Ciuchini:1993ks}. Earlier analyses \cite{Grinstein:vj}--\cite{Misiak:bc} contained
additional approximations or were not fully correct. The results of
ref.~\cite{Ciuchini:1993ks} were subsequently confirmed in
refs.~\cite{Misiak:bc-erratum}--\cite{Chetyrkin:1996vx}.

As pointed out in refs.~\cite{Ali:1993ct,Buras:xp}, the LO expression
for $\Gamma[b \to X_s^{\rm parton} \gamma]$ suffers from large
($\sim\;\pm 25\%$) renormalization scale uncertainties. Therefore,
matching the experimental accuracy of eq.~(\ref{exp}) requires
performing a complete NLO QCD calculation. This goal has been achieved
in a joint effort of many groups:
\begin{itemize}
\item
Two-loop ${\cal O}(\alpha_s)$ corrections to the matching
conditions $C_7(\mu_0)$ and $C_8(\mu_0)$ were first calculated in
ref.~\cite{Adel:1993ah} and subsequently confirmed by several groups
\cite{Greub:1997hf}--\cite{Bobeth:1999ww}.
\item
Two-loop mixing and one-loop matching for the four-quark operators
$Q_1,...,Q_6$ were found in refs.~\cite{Altarelli:1980fi}--\cite{Ciuchini:1992tj}.  In
ref.~\cite{Chetyrkin:1997gb}, these results were confirmed by recalculation in a
different operator basis that is more suitable for $b \to s \gamma$
analyses.
\item
Two-loop mixing in the sector $(Q_7,Q_8)$ was calculated in ref.~\cite{Misiak:1994zw}.
These results have been recently confirmed \cite{GGHunp}.
\item
Three-loop mixing between the sectors $(Q_1,...,Q_6)$ and 
$(Q_7,Q_8)$ was evaluated in ref.~\cite{Chetyrkin:1996vx}.
It is currently being verified by another group \cite{GGHunp}.
\item
The leading-order matrix elements 
$\langle s\gamma g|Q_i| b\rangle$ and the one-loop matrix element
 $\langle s\gamma |Q_7| b\rangle$ 
were calculated in refs.~\cite{Ali:1990tj,Pott:1995if}. Some of them were
confirmed in ref.~\cite{Ligeti:1999ea} where certain BLM corrections were
included, too.
\item
Two-loop calculation of the matrix element 
$\langle s\gamma |Q_{1,2}| b\rangle$ 
was presented in ref.~\cite{Greub:1996jd}. It has been recently
verified and extended to the full basis of four-quark operators
\cite{Buras:2002tp,Buras:2001mq}. The one-loop matrix element $\langle s\gamma |Q_8| b\rangle$ 
has been found in refs.~\cite{Greub:1996jd,Buras:2002tp}, too.
\end{itemize}
It should be emphasized that all these ingredients enter not only the
analysis of ${\bar B}\to X_s\gamma$ in the SM but are also necessary
in extensions of this model. The corrections to the Wilson
coefficients of the operators $Q_7$ and $Q_8$ are also relevant for
$\bar{B} \to X_s l^+l^-$.

\newsection{Electroweak corrections}
\label{sec:ew}

The study of electroweak corrections begins with searching for terms
that might be enhanced by large logarithms. Czarnecki and Marciano
\cite{Czarnecki:1998tn} pointed out that large logarithms ~$\ln(m_b^2/m_e^2)$ \linebreak 
are absent when ~$\alpha_{\rm em}^{\rm on\;shell}$~ is used in the overall
normalization of ~$\Gamma[b \to X_s^{\rm parton} \gamma]$.

Another type of large logarithm that might enhance some of the
electroweak corrections is ~$\ln(m_W^2/m_b^2)$,~ i.e. the same
logarithm that is responsible for the huge QCD enhancement of the $b
\to s \gamma$ amplitude. Once ~$[1-\al(\mu_0)/\al(\mu_b)] \sim 0.4$~
is treated as a quantity of order unity, the considered electroweak
correction is formally of order ${\cal O}(\alpha_{\rm em}/\al)$, so it
might be numerically relevant, given the accuracy in
eq.~(\ref{exp}). However, as demonstrated in
refs.~\cite{Kagan:1998ym,Czarnecki:1998tn,Baranowski:1999tq} through
explicit calculations, it turns out to be negligible ($\sim -0.7\%$).

The articles \cite{Gambino:2000fz} contain results for the complete electroweak
corrections to the matching conditions $C_i(\mu_0)$.  Some of them are
proportional to $\alpha_{\rm em}(M_Z)/\sin^2\theta_W \simeq 0.034$.
Their effect on ~$\Gamma[b \to X_s^{\rm parton} \gamma]$~ amounts\footnote{
This number includes QED corrections to the matrix elements of $Q_{1,2,7}$, too.}
to ~$-1.5\%$ for $M_{\rm Higgs} = 115\;$GeV, and diminishes with
increasing $M_{\rm Higgs}$. The authors of ref.~\cite{Gambino:2000fz} resolved
the numerical discrepancy between refs.~\cite{Strumia:1998bj} and \cite{Czarnecki:1998tn}
in favor of the latter.

The only electroweak ${\cal O}(\alpha_{\rm em})$ corrections that
remain unknown at present are enhanced neither by large logarithms nor
by $1/\sin^2\theta_W$.  Thus, we can be practically certain about
their irrelevance.

\newsection{Non-perturbative effects}
\label{sec:nonpert}

The LO contribution to $\Gamma[b \to X_s^{\rm parton} \gamma]$ is
given by the tree-level matrix element of the $Q_7$ operator.\footnote{
In dimensional regularization, one-loop matrix elements of
$Q_3,...,Q_6$ may give LO contributions, too. However, they can be
absorbed into the tree level matrix element of $Q_7$ with a redefined
Wilson coefficient \cite{Buras:xp,Misiak:dj}.}
Let us temporarily assume that this operator is the only one in the
effective Lagrangian (\ref{Leff}), and denote the corresponding
contribution to the hadronic width by
$\Gamma[\bar{B} \to X_s \gamma]^{(Q_7\;{\rm only})}$. 

In analogy to the analyses \cite{Chay:1990da,Manohar:1993qn} of the inclusive
semileptonic decay $\bar{B} \to X_u e \bar{\nu}$, one can apply the
Operator Product Expansion (OPE) and Heavy Quark Effective Theory (HQET) 
to show that
\be \label{nonpert1}
\Gamma[\bar{B} \to X_s \gamma]^{(Q_7\;{\rm only})}
= \Gamma[b \to X_s^{\rm parton} \gamma]^{(Q_7\;{\rm only})}
 \left[ 1 + a_1 \f{\lambda_1}{m_b^2} + a_2 \f{\lambda_2}{m_b^2} 
   + {\cal O}\left(\f{\Lambda^3_{\scs \rm QCD}}{m_b^3}\right)\right].
\ee
Here, $\lambda_{1,2} \sim \Lambda_{QCD}^2$ are the standard HQET
parameters.  The value of $\lambda_2 \simeq 0.12\;{\rm GeV}^2$ is
known from the measured $B$--$B^*$ mass difference. The value of
$\lambda_1 = -(0.27\pm0.10\pm0.04)\;{\rm GeV}^2$ has been determined
in ref.~\cite{Hoang:1998ng} from the observed semileptonic $B$-decay
spectra (see ref.~\cite{Bauer:2002kk} for more recent determinations).
The coefficients $a_1$ and $a_2$ can be calculated within perturbation
theory,\footnote{
The same refers to similar coefficients at higher orders in the 
($\Lambda_{\scs \rm QCD}/m_b$)-expansion.}
which yields \cite{Falk:1993dh,Bigi:1992ne}
\be \label{nonpert2}
a_1 ~=~ \f{1}{2} ~+~ {\cal O}(\al(m_b))
\hspace{2cm} {\rm and} \hspace{2cm} 
a_2 ~=~ -\f{9}{2} ~+~ {\cal O}(\al(m_b)).
\ee
The resulting ${\cal O}(\Lambda^2_{\scs \rm QCD}/m_b^2)$
non-perturbative correction on the r.h.s. of eq.~(\ref{nonpert1}) amounts
to around ~$-3\%$.

\begin{figure}[t]
\begin{center}
\includegraphics[width=7cm,angle=0]{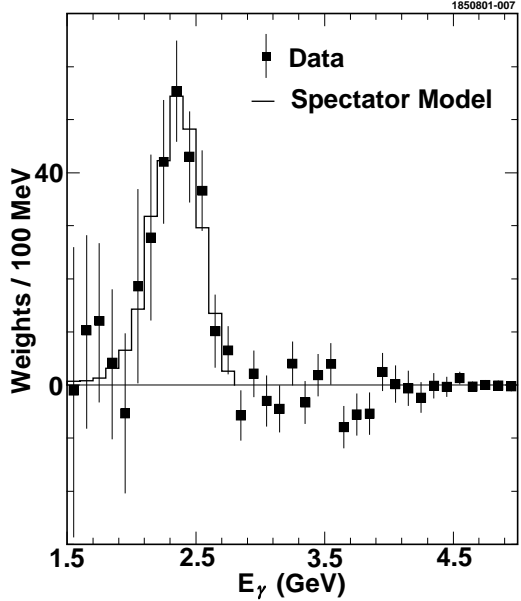}\\[-3mm]
\caption{The $\bar{B} \to X_s \gamma$ photon spectrum observed by CLEO \cite{Chen:2001fj}.}
\label{fig:exp.spectrum}
\end{center}
\vspace*{-6mm}
\end{figure}

The relation (\ref{nonpert1}) still holds when a lower cutoff $E_0$ is
imposed on the photon energy in the ${\bar B}$-meson rest frame,
provided $E_0$ is not too close to the endpoint
$E^{\rm max} = \f{m_B^2-m_{K^*}^2}{2m_B} \simeq 2.6\;{\rm GeV}$. \linebreak
Acceptable values of $E_0$ must correspond to much larger than
$\Lambda_{\scs \rm QCD}$ invariant masses of the recoiling hadronic
state $X_s$. Fig.~3 in ref.~\cite{Kagan:1998ym} suggests that $E_0 = 1.6\;{\rm
GeV}$ is sufficiently low. More than 95\% of the total $\Gamma[b \to
X_s^{\rm parton} \gamma]$ originates from a peak that lays above such
a cutoff.\footnote{
Consequently, $E_0$-dependence in eq.~(\ref{nonpert2}) can be safely neglected.}
This peak is now clearly seen in the $\bar{B} \to X_s \gamma$ spectrum
observed by CLEO (fig.~\ref{fig:exp.spectrum}).  Its position
corresponds to the photon energy in the leading two-body decay $b \to
s \gamma$.

There is neither experimental nor theoretical need to consider photons
below 1.6~GeV. They are practically unobservable at the inclusive
level, because of the overwhelming $b\to c$ background. On the
theoretical side, keeping not too small $E_0$ facilitates the
discussion of non-perturbative effects due to operators other than
$Q_7$. Of course, we have to admit that 1.6~GeV is chosen arbitrarily.
It could almost equivalently be 1.5 or 1.7~GeV. However, going up to
the current CLEO cutoff of 2.0~GeV would increase uncertainties on the
theoretical side. Data-driven extrapolation from the experimental
cutoff to the theoretically preferred one is the right choice to make
at present.

The discussion of non-perturbative effects becomes much more complex
when we take into account operators other than $Q_7$. It is no longer
possible to apply OPE in analogy to $\bar{B} \to X_u e \bar{\nu}$,
because the $b$-quark annihilation and the photon emission may now be
separated in space-time by more than $\Lambda_{\scs \rm QCD}^{-1}$.

The contribution of $Q_8$ to $\Gamma[\bar{B} \to X_s \gamma]$ has been
analyzed in ref.~\cite{Kapustin:1995fk} with the help of fragmentation
functions. Important non-perturbative effects have been found for low
$E_{\gamma}$ only, i.e. much below $E_0=1.6\;$GeV. Thus, with our
cutoff, a reliable approximation is given by the perturbative
contribution to $\Gamma[b \to X_s^{\rm parton} \gamma]$ from the
matrix elements of $Q_8$.  The accuracy of this approximation does not
need to be known precisely, because the perturbative contribution of
$Q_8$ is smaller than 3\%.

Similar conclusions can be drawn for the operators $(\bar{s} \Gamma
b)(\bar{q} \Gamma' q)$, where $q=u,d,s$. They are present inside
$Q_3,..., Q_6$.  Perturbative effects of their matrix elements are
even smaller than that of $Q_8$. As far as non-perturbative effects
are concerned, one might worry about production of virtual vector
mesons that convert to a real photon. However, creation of such
transverse mesons is impossible in the factorization approximation
because $Q_3,..., Q_6$ contain no ~$\bar{q} \sigma_{\mu\nu} q$~
currents.  Deviations from the factorization approximation are
suppressed either by $\al(m_b)$ or by $\Lambda_{\scs \rm QCD}/m_b$
\cite{Beneke:1999br}. This is sufficient to make them negligible here, given
the smallness of ~$|C_{3, ..., 6}(\mu_b)| < 0.07$,~ as compared to
~$|C_{1,2,7,8}(\mu_b)| \simeq ( 0.5,\; 1,\; 0.3,\; 0.15 )$.

The operators $Q_3,..., Q_6$ contain $(\bar{s} \Gamma b)(\bar{b}
\Gamma' b)$ terms, too. The $b$-quark loops are localized at distances
much smaller $\Lambda_{\scs \rm QCD}^{-1}$ in space-time. Thus, they
can undergo the same treatment as $Q_7$, as far as non-perturbative
effects are concerned.  Since their perturbative contributions are
minor, the non-perturbative ones are totally negligible.

Charm quark loops are the most difficult to analyze. Factorization 
is not sufficient here because $2 m_c/m_b$ is not a small number.
Moreover, non-factorizable contributions may be numerically important
because the Wilson coefficients $C_1$ and $C_2$ are not small at all.

Let us begin with tracing down possible contributions from
intermediate real $c\bar{c}$ states. Our cutoff $E_0=1.6\;$GeV implies
that the invariant mass of the final $X_s$ state is smaller than
$m_{\eta_c}+m_K$. Consequently, real $c\bar{c}$ states might occur only
{\em before} the photon emission, i.e. in a cascade decay: $\bar{B} \to
Y_{c\bar{c}} X_s^{(1)}$ followed by $Y_{c\bar{c}} \to X^{(2)}\gamma$.

The importance of such processes can be tested in the case
$Y_{c\bar{c}} = \psi$, because separate experimental data on both
(inclusive) components of the cascade decay are available. For low
$E_0$, the resulting branching ratio of the intermediate $\psi$
contribution is larger than the one in eq.~(\ref{exp}). It gets
reduced to (a few)$\times 10^{-5}$ for $E_0=1.6\;$GeV, and becomes
negligible for $E_0=2.0\;$GeV \cite{KMu}.

\begin{figure}[h]
\begin{center}
\includegraphics[width=25mm,angle=0]{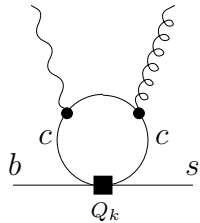}
\end{center}
\vspace{-17mm}
\hspace*{75mm} $c$ \hspace{11mm}  $c$\\[-1mm]
\hspace*{71mm} $b$ \hspace{19mm} $s$\\[2mm]
\hspace*{82mm} $^{Q_k}$\\[-12mm]
\begin{center}
\caption{\sf Charm loop contribution to $b \to s \gamma g$.}
\label{fig:p2brems}
\end{center}
\vspace*{-1cm}
\end{figure}

The models used by CLEO \cite{Chen:2001fj,Thorndike:2002rt} to
extrapolate from 2.0~GeV to lower cutoffs do not include the
intermediate $\psi$ contribution. They are based on perturbative
calculations, in which the only diagram (fig.~\ref{fig:p2brems}) that
might correspond to this contribution affects $\Gamma[b \to X_s^{\rm
parton} \gamma]$ by 1.7\% only (for $E_0=1.6\;$GeV). Consequently, the
procedure applied by CLEO is consistent with treating the intermediate
$\psi$ contribution as background.

Identical arguments work for $\psi'$. Higher $c\bar{c}$ states might
produce higher energy photons. However, radiative charm annihilation
processes in all the $c\bar{c}$ states except $\psi$ and $\psi'$ have
negligible branching ratios. Thus, it does not really matter whether
we consider their contributions as background or not. Whatever
decision is made, its effect is expected to be less than the 1.7\%
perturbative contribution from the diagram in fig.~\ref{fig:p2brems}.

Having discussed the real intermediate $c\bar{c}$ states, we proceed
to the virtual ones. Neither infrared nor collinear singularities
occur in the perturbative contributions of $c\bar{c}$ loops to
\linebreak $\Gamma[b \to X_s^{\rm parton} \gamma]$ at NLO. Thus,
according to the common wisdom, one expects that these perturbative
results give reasonable estimates to the corresponding contributions
to $\Gamma[\bar{B} \to X_s \gamma]$, up to corrections of order 
${\cal O}(\Lambda_{\scs \rm QCD}/m_{c,b})$.

\begin{figure}[h]
\begin{center}
\includegraphics[width=25mm,angle=0]{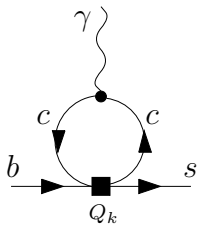}
\end{center}
\vspace{-33mm}
\hspace*{8cm} $\gamma$\\[8mm]
\hspace*{75mm} $c$ \hspace{1cm}  $c$\\[2mm]
\hspace*{71mm} $b$ \hspace{19mm} $s$\\[2mm]
\hspace*{82mm} $^{Q_k}$\\[-11mm]
\begin{center}
\caption{\sf One-loop matrix element that vanishes for the on-shell photon.}
\label{fig:mel1}
\end{center}
\vspace*{-1cm}
\end{figure}

The actual situation is somewhat more complicated, because the leading
one-loop diagram (fig.~\ref{fig:mel1}) vanishes for the on-shell
photon. However, it becomes non-vanishing when a soft gluon is
attached to the $c$-quark loop. Such a gluon may originate from the
decaying $\bar{B}$ meson. Thus, one finds a non-perturbative effect
\cite{Voloshin:1996gw,Khodjamirian:1997tg} that is not approximated in
any sense by the corresponding perturbative null. Fortunately, it can
be expressed within HQET in terms of a series
\be \label{nonpert3}
\f{\Delta \Gamma[\bar{B} \to X_s \gamma]}{\Gamma[\bar{B} \to X_s \gamma]}
~=~ \f{\lambda_2}{m_c^2} \; \sum_{n=0}^{\infty} 
b_n \left(\f{m_b \Lambda_{\scs \rm QCD}}{m_c^2} \right)^n,
\ee
in which the $n\geq 1$ terms are likely to be negligible, because the
coefficients $b_n$ decrease rapidly with $n$
\cite{Ligeti:1997tc,Grant:1997ec}.  The calculable leading ${\cal
O}(\Lambda^2_{\scs \rm QCD}/m_c^2)$ term enhances the decay width by
around 2.5\% \cite{Buchalla:1997ky}.

The perturbative ${\cal O}(\al)$ results described in
section~\ref{sec:perturb} include non-vanishing two-loop diagrams with
$c\bar{c}$ loops, e.g. the ones obtained by adding a virtual gluon to
the diagram in fig.~\ref{fig:mel1}. The corresponding non-perturbative
effects are expected to be suppressed by both $\al(m_b)$ and
$\Lambda_{\scs \rm QCD}/m_{c,b}$. Thus, at the first glance, they
might seem irrelevant. However, it remains an open question whether
their suppression is numerically sufficient. No quantitative estimates
of such non-perturbative effects have been performed so far. We shall
discuss this issue in more detail at the end of the next section.

\newpage
\newsection{Phenomenological discussion}
\label{sec:phenom}

In the present section, we shall discuss the two main uncertainties in
the present-day SM prediction for $\bar{B} \to X_s \gamma$.  The
analysis of ref.~\cite{Gambino:2001ew} will be largely followed.

The prediction (\ref{thSM}) is obtained from the formula
\mathindent0cm
\bea \label{main}
{\rm BR}[\bar{B} \to X_s \gamma ~ (E_{\gamma} > E_0)] && \nonumber \\[2mm]
&& \hspace{-25mm} = {\rm BR}[\bar{B} \to X_{c\,} e \bar{\nu}]_{\rm exp}
\left( \f{\Gamma[\bar{B} \to X_u e \bar{\nu}]} 
         {\Gamma[\bar{B} \to X_{c\,} e \bar{\nu}]} \right)_{\rm th} 
\left( \f{\Gamma[\bar{B} \to X_s \gamma ~ (E_{\gamma} > E_0)]}
         {\Gamma[\bar{B} \to X_u e \bar{\nu}]} \right)_{\rm th},
\eea
\ \\[-2mm] 
in which the following substitutions are made\\[-2mm]
\bea 
\left( \f{\Gamma[\bar{B} \to X_s \gamma ~ (E_{\gamma}\! >\! E_0)]} 
         {\Gamma[\bar{B} \to X_u e \bar{\nu}]} \right)_{\rm th} \!\!
&\simeq& \label{bsg.ov.bu}
\left( \f{\Gamma[ b \to X_s^{\rm parton} \gamma ~ (E_{\gamma}\! >\! E_0)]} 
         {\Gamma[b \to X_u^{\rm parton} e \bar{\nu}]} 
\right)_{\rm \! NLO} \! + \left( \begin{array}{c} \mbox{\small non-perturbative} \\ 
                \mbox{\small corrections (\ref{nonpert3})} \end{array} \right)\! ,\\[4mm]
\left( \f{\Gamma[\bar{B} \to X_u e \bar{\nu}]} 
         {\Gamma[\bar{B} \to X_{c\,} e \bar{\nu}]} \right)_{\rm th} \!\!
&\simeq& \label{bu.ov.bc}
\left( \f{\Gamma[b \to X_u^{\rm parton} e \bar{\nu}]} 
         {\Gamma[b \to X_c^{\rm parton} e \bar{\nu}]} 
\right)_{\rm \! NNLO} \! + \left( \begin{array}{c} 
\mbox{\small known}~\; {\cal O}\!\left(\lambda_2/m_b^2 \right) \\ 
\mbox{\small corrections} \end{array} \right).
\eea
\mathindent1cm
\ \\[-2mm] 
Such ratios are introduced in order to minimize uncertainties in
eq.~(\ref{main}) that originate from the CKM angles and the overall
factors of $m_b^5$. The use of $b\to u$ transitions is motivated by
the fact that eq.~(\ref{bu.ov.bc}) is known at the NNLO, while
convergence of the perturbation series and non-perturbative effects
are more easily controlled in eq.~(\ref{bsg.ov.bu}) than in
$\Gamma[\bar{B} \to X_s \gamma]/\Gamma[\bar{B} \to X_{c\,} e
\bar{\nu}]$.  The ${\cal O}(\Lambda^2_{\scs \rm QCD}/m_b^2)$ terms
from eq.~(\ref{nonpert1}) have canceled in the ratio (\ref{bsg.ov.bu})
with the analogous corrections to $\Gamma[\bar{B} \to X_u e
\bar{\nu}]$. On the other hand, once the quark masses are expressed in
terms of the hadronic ones, the ratio (\ref{bu.ov.bc}) depends on both
$\lambda_1$ and $\lambda_2$. 

\begin{figure}[h]
\begin{center}
\includegraphics[width=14cm,angle=0]{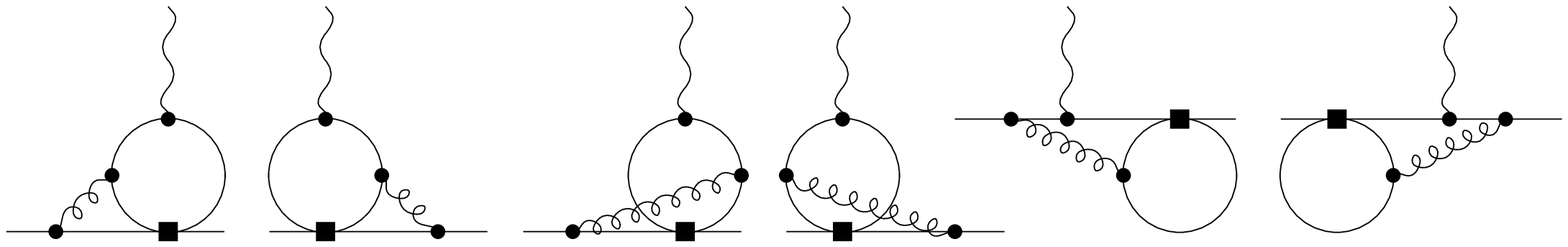}
\caption{\sf Charm loop contributions to the matrix elements of four-quark operators.}
\label{fig:ghw.diag}
\end{center}
\ \\[-36mm]
\hspace*{10cm}$b$\hspace{22mm}$s$\hspace{4mm}$b$\hspace{21mm}$s$\\[-3mm]
\hspace*{24mm}$c$\hspace{13mm}$c$\hspace{3cm}$c$\hspace{13mm}$c$\\[3mm]
\hspace*{113.5mm}$c$\hspace{12.5mm}$c$\\
\hspace*{15mm}$b$\hspace{16mm}$s$\hspace{5mm}$b$\hspace{15mm}$s$
  \hspace{3mm}$b$\hspace{14mm}$s$\hspace{5mm}$b$\hspace{16mm}$s$\\[-1mm]
\end{figure}
 
The main uncertainty in the perturbative ratio on the r.h.s. of
eq.~(\ref{bsg.ov.bu}) originates from the two-loop diagrams with charm
quarks presented in fig.~\ref{fig:ghw.diag}.  Such diagrams are the
only source of $m_c$-dependence of this ratio. Since the higher-order
(NNLO) QCD corrections are unknown, the renormalization scheme for
$m_c$ remains arbitrary, at least within a certain class of
``reasonable'' schemes that do not artificially enhance the unknown
corrections. As argued in ref.~\cite{Gambino:2001ew}, the uncertainty in
eq.~(\ref{thSM}) stemming from this scheme-dependence can be accounted
for by setting $m_c/m_b = m_c(\mu)^{\overline{\rm MS}}/m_b^{1S}$ in
the two-loop diagrams,\footnote{
Here, $m_b^{1S}$ stands for the $b$-quark mass in the so-called
``1S-scheme'' \cite{Hoang:1998ng}. It is defined as half of the perturbative
contribution to the $\Upsilon$ mass.}
and varying the scale $\mu$ between $m_c$ and $m_b$. Such a variation
is the dominant source of the error in eq.~(\ref{thSM}).

One could remove the considered uncertainty by calculating three-loop
diagrams obtainable from fig.~\ref{fig:ghw.diag} by adding one more
virtual gluon. UV-divergent parts of such diagrams have been already
found in the process of calculating the NLO anomalous dimensions
\cite{Chetyrkin:1996vx}. Evaluating the finite parts would constitute
an extremely tedious task, though not totally impossible, if numerical
integration was applied. Finding the remaining NNLO corrections would
be relatively simpler, given that fully automatized analytical methods
are now available
\cite{vanRitbergen:1997va,Chetyrkin:1996hq,Steinhauser:2000ry}.

However, before undertaking such an ambitious task, one should make
sure that all the non-perturbative effects are really under control.
The main worry are the doubly-suppressed corrections mentioned in the
last paragraph of section~\ref{sec:nonpert}. So far, they have been
neither estimated nor included in the theoretical error. They are
related to {\em precisely} the same two-loop diagrams with charm
quarks (fig.~\ref{fig:ghw.diag}). Numerical importance of non-local
parts of those diagrams\footnote{
By non-local we mean those parts that cannot be removed off-shell by
finite local counterterms.}
can be illustrated by the fact that the r.h.s. of eq.~(\ref{thSM})
changes by 35\% when $m_c$ is shifted from the original value of
$\;0.22\, m_b\;$ to the threshold for charm pair production $m_c =
\f{1}{2} m_b$. A ~$\Lambda_{\scs \rm QCD}/m_b$-suppressed
non-perturbative effect on the top of such a large perturbative
contribution might not be negligible. Unfortunately, no systematic
methods have yet been developed for calculating corrections of this
type.

\newsection{Conclusions}
\label{sec:concl}

In the present paper, we have summarized the existing calculations of
perturbative and non-perturbative contributions to the inclusive weak
radiative $\bar{B}$ meson decay. We have pointed out that both the
main perturbative uncertainty and the most worrisome non-perturbative
effects have their origin in the fact that non-local charm quark loop
contributions are particularly large. Removing the perturbative
uncertainty due to $m_c$-dependence would be extremely tedious, but
not totally impossible. However, developing a method for
systematically estimating the related non-perturbative effects is
desirable in advance.

The present agreement at the $\sim 10\%$ level between the
experimental (\ref{exp}) and theoretical (\ref{thSM}) determinations
of ${\rm BR}[\bar{B} \to X_s \gamma]$ implies that clear signatures of
new physics in this observable are not likely to be found in the
foreseeable future. The importance of improving the accuracy on both
the experimental and theoretical sides follows from the need for
strengthening the $b \to s \gamma$ {\em constraints} on beyond-SM theories.
Such constraints are likely to be crucial in identifying the origin of
new physics effects that we expect to encounter in the LHC era.

\newsection{Acknowledgements}

We would like to thank K.~Baranowski, C.~Bobeth, K.G.~Chetyrkin,
A.~Czarnecki, P.~Gambino, M.~Jamin, A.~Khodjamirian, A.~Kwiatkowski,
M.E.~Lautenbacher, M.~M\"unz, S.~Pokorski, N.~Pott, J.~Urban and
P.H.~Weisz for fruitful collaboration in the calculations that were
relevant for determining the SM prediction for $b \to s
\gamma$. A.J.B. is grateful to the Max Planck Institute for Physics
for financial support during the Planck02 Symposium in Kazimierz.  His
work presented here has been supported in part by the German
Bundesministerium f\"ur Bildung und Forschung under the contract
05HT1WOA3 and the DFG Project Bu.  706/1-1. M.M. was supported in part by
the Polish Committee for Scientific Research under the grant
2~P03B~121~20.

\setlength {\baselineskip}{0mm}
 
\end{document}